      [1999/12/01 v1.4c Il Nuovo Cimento]
\documentclass{cimento}
\usepackage{graphicx}  % got figures? uncomment this

\title{Polarons and Bipolarons in Holstein and Holstein $t-J$ models by
Dynamical Mean Field Theory}
\author{S.
Ciuchi\from{ins:AQ},M.Capone\from{ins:RM},E.Cappelluti\from{ins:RM} and G. Sangiovanni\from{ins:Stutt}}
\instlist{
\inst{ins:AQ} Dipartimento di Fisica
Universit\`a dell'Aquila,
via Vetoio, I-67010 Coppito-L'Aquila, Italy
\inst{ins:RM} 
``Istituto dei Sistemi Complessi'', CNR-INFM,
v. dei Taurini 19, 00185 Roma, Italy,
\inst{ins:Stutt} Max-Planck Institut f\"ur Festk\"orperforschung, Heisenbergstr. 1, D-70569 Stuttgart, Germany
}
\PACSes{\PACSit{71.38.-k}\PACSit{71.30.+h}\PACSit{71.38.Ht}\PACSit{71.10.Fd}}

\newcommand{\beq}{\begin{equation}}
\newcommand{\eeq}{\end{equation}}
\newcommand{\beqa}{\begin{eqnarray}}
\newcommand{\eeqa}{\end{eqnarray}}

\newcommand{\D}{\Delta}
\newcommand{\e}{\epsilon}
\newcommand{\sx}{\sigma_x}
\newcommand{\sz}{\sigma_z}
\newcommand{\s}{\alpha}		% the phonon spin
\renewcommand{\l}{\eta}		% the coefficient which was formerly called \lambda in spi.pdf
\renewcommand{\o}{\omega}
\newcommand{\ad}{\gamma}	% the symbol for adiabatic parameter w0/t

\newcommand{\kvec}{{k}}	% to be properly referred as state index

\begin{document}

\maketitle

\section{Introduction}
\label{sec:intro}

In system with strong electron-phonon (e-ph) interaction, the carriers lose mobility, 
ultimately  acquiring
polaronic character. A polaron is a state in which the phonon and electron
degrees of freedom are strongly entangled, and the presence of an electron is
associated to a finite lattice distortion, which in  turn binds the electron
leading to the so-called self-trapping effect. Polarons also tend to create
bound pairs, called bipolarons of course the presence of Coulomb repulsion
destabilize bi-polarons in favor of a pure polaronic state at finite densities
\cite{HewsonVarenna,KastellaniVarenna}. Typical signatures of polarons are seen 
in multi-peaked photoemission spectra\cite{EgamiVarenna} and
transport measurements, where an  activated behavior with a characteristic   
energy given by the polaronic binding energy is observed.   The polaronic peak
found in the mid infrared measurements of optical conductivity 
\cite{CalvaniVarenna} may also not only detect the polaronic binding energy 
\cite{Fehske-Akw} but
also other subtle polaronic transitions at very low energy \cite{frat1}.
Another less classical indication of polaronic formation comes from the analysis
of lattice displacements associated to the excess charge as obtained by the
distribution of distances between atoms\cite{EgamiVarenna}.

A joint analysis of both spectral \cite{Fehske-Akw} and local lattice distortions
can  disentangle various kind of behavior in polaronic systems.
In fact we notice that neglecting the repulsive interaction a gapped 
pair state can be formed
even without a significant associated polarization. In this case bipolarons are
expected to be a relatively mobile particle.
% and 
%thermal activation  will be not associated to the single polaron formation
%\cite{frat2}  but rather to the existence of an insulating state which is due
%to pairing correlation effects.

The aim of this work is to provide a thorough analysis of polaronic spectral
properties in both single and multi-polaron cases.  The backbone of our
presentation is the DMFT which is introduced and briefly discussed in general
in section \ref{sec:DMFT}.

In the single polaron case we review the exact solution of the Holstein model 
\cite{sumi,depolarone} (section \ref{sec:HolsteinAndHtJ}) and we  present some
new results for the Holstein $t-J$ model comparing our results with those of
Mishchenko \cite{MishchenkoVarenna}. At large polaronic densities we use both
Exact Diagonalization (ED) and Quantum Monte Carlo (QMC) techniques to solve the 
DMFT equations for the Holstein model respectively at
the $T=0$ (sec. \ref{sec:HHFED}) and  $T>0$ (section \ref{sec:HHFQMC}).
In this case we compare spinless and spinful fermions cases
and we discuss in detail the role of the adiabatic ratio. In this way the
properties of a pure polaronic state can be disentangled from those of
bipolaronic state.   We compare also numerical solutions with analytic
approximate schemes based on  Born Oppenheimer (\ref{sec:HHFEDad})  and 
Lang-Firsov canonical
transformation (\ref{sec:HHFEDantiad})  respectively in adiabatic and
antiadiabatic regime.

% We present some analytical as well as numerical  solution for
%Holstein and Holstein-$t-J$ models both a zero and finite temperatures. DMFT 
%does not rely on any small parameter assumption. Such results allow us to
%identify the virtues and defects of various analytic approximated schemes. 

%The 
%In
%particular, in the multipolaronic case, we discuss in detail a Born-Oppenheimer
%(BO) scheme which is based on the adiabatic limit,  and discuss an
%antiadiabatic approach slightly different from that universally accepted in
%dealing with polaronic systems \cite{Lang-Firsov}. In particular the BO scheme
%which will be presented here is suitable to be applied to {\it both} the weak
%and the strong coupling regime.

\section{The DMFT method}
\label{sec:DMFT}

\subsection{Introduction}
\label{sec:naive}
Dynamical Mean Field Theory is a non-perturbative technique originally
developed as the exact solution of a interacting electron problem on an infinite
dimensional lattice \cite{BrandtMielsch,Muller-Hartmann}. 
A comprehensive review can be found in ref. \cite{DMFTreview} now let us sketch some key
point to understand the developments presented in following sections.
Let a consider a general tight-binding problem on a lattice with coordination
number $z$
\beq
\label{hamiltonian0}
H = -\frac{t}{\sqrt{z}}\sum_{\langle ij \rangle \sigma} c^\dagger_i c_j + \sum_i V[c_i,c^\dagger_i]
\eeq
where $c_i$ are fermionic annihilation operator acting on site $i$ (spin index
omitted for simplicity) $V$ is a local (on-site) potential. 
The scaling of hopping $t$ is such as the limit $z\rightarrow \infty$ give a non
trivial result.
Mean field theory turns out to be exact in infinite dimensions i.e. when the
number of nearest neighbor diverges, therefore we can
replace the effect of hopping on neighboring sites as {\it cavity field} $\eta$
$$
\eta_i= \frac{t}{\sqrt{z}}\sum^{(i)}_j c_j 
$$
with the sum running on nearest neighbor of site $i$. In terms of the internal
fields $\eta$, Hamiltonian can be written formally as a sum of single site operators as
\beq
H = --\sum_i \eta^\dagger_i c_i-\sum_i c^\dagger_i\eta_i  + 
\sum_i V[c_i,c^\dagger_i]
\eeq
For fermions $\eta$ obeys  anticommutation relations 
$$
\left [ \eta_i \eta^\dagger_j\right ] = t\delta_{i,j} \\
\left [ \eta_i c^\dagger_j\right ] = \frac{t}{\sqrt{z}}\delta_{i,j}.
$$  
On a more formal ground it can be demonstrated that the cavity field is a
Gaussian Grassmann field which is therefore determined solely by its correlation
function\cite{DMFTreview}. 

\subsection{Single impurity action}
\label{sec:ImpurityAction}
The previous arguments can be more formally developed using a path-integral formalism.
In analogy with classical mean-field\cite{DMFTreview}, the fermions of all sites
but one  (namely $0$) are integrated out leading to a 
single-site partition function 
\begin{equation}
\label{Zpart}
Z = \int \Pi_i \mathcal{D} \psi^{\dagger} \mathcal{D} \psi  \exp(-S)
\end{equation}
where now $\psi$ and $\psi^\dagger$ are Grassmann anticommuting  
eigenvalues of the creation/destruction operators of site $0$\cite{Negele}. 
The action $S$ is given by
\begin{equation}\label{Simpurity}
S = -\int_0^\beta d\tau \int_0^\beta d\tau' 
 \psi^\dagger(\tau) \mathcal{G}^{-1}_0(\tau-\tau') \psi(\tau') + 
 \int_0^\beta d\tau V[\psi(\tau),\psi^\dagger(\tau)]
\end{equation}
where the correlator $\mathcal{G}^{-1}_0(\tau-\tau')$ is 
\beq
\label{selfcons0}
\mathcal{G}^{-1}_0(\tau-\tau')= \partial_\tau
\delta(\tau-\tau')+\langle\eta_i(\tau)\eta_i(\tau')\rangle
\eeq
which is independent on $i$ due to translation invariance.
The action (\ref{Simpurity}) 
depends parametrically on the environment through the correlator $\mathcal{G}_0$. 
(\ref{Simpurity}) is indeed the action of  a {\it single} impurity embedded in a
medium whose properties are related to the original lattice via  the 
self-consistency equation 
(\ref{selfcons0}).
To be more concrete let us consider an infinite coordination Bethe lattice of
half-bandwidth $D$, for which the Green's function of $\eta$ is proportional to 
the local Green's function $G_{j,j}$ \cite{DMFTreview}
$$
\langle\eta_i(\tau)\eta_i(\tau')\rangle=\frac{D^2}{4z}\sum_{j} G_{j,j}
$$
then Eq. (\ref{selfcons0}) reads 
$\mathcal{G}^{-1}_0(\tau)=-\partial_\tau+(D^2/4) G(\tau)$,
or in frequency domain
\begin{equation}
\label{selfcons2}
\mathcal{G}^{-1}_0(i\omega_n)=i\omega_n+(D^2/4) G(i\omega_n).
\end{equation}
Eqs (\ref{Zpart}) and (\ref{Simpurity}), together with the self-consistency condition
(\ref{selfcons2}) form a closed set of mean-field equations.

\subsection{Single impurity Hamiltonian}
\label{sec:ImpurityHamiltonian}
A Hamiltonian formalism can be also developed when suitably defined fermion
fields are introduced to get required gaussian cavity field. Eq.
(\ref{Simpurity}) can be obtained by integrating out  auxiliary 
fermionic fields $c_k$ of an Anderson Impurity Hamiltonian (AIM)
\beq
\label{AndersonImpurity}
H_{AIM} = \sum_k V_k (f^\dagger c_k+c_k f^\dagger)+\sum_k E_k c^\dagger c_k+
V[f,f^\dagger]
\eeq
levels $E_k$ and hybridization constants $V_k$ must be chosen to give the
appropriate cavity field. In  the Bethe lattice case
\begin{equation}
\label{eq:self-consAIM}
\frac{D^2}{4}G(i\omega_n) = \sum_k \frac{V_k^2}{i\omega_n - E_k}.
\end{equation}
where in the l.h.s.  we read the local {\it
lattice} propagator.
Eq. (\ref{eq:self-consAIM}) becomes the self-consistency
condition which determines the appropriate AIM parameters $E_k$ and $V_k$.

\section{Holstein model in infinite dimensions}
\label{sec:Holstein}
The Holstein molecular crystal model is the paradigmatic model for small
polarons. Its Hamiltonian reads

\begin{eqnarray}
\label{eq:themodel}
H &=& -t\sum_{\langle i,j\rangle,\sigma} (c^{\dagger}_{i,\sigma} c_{j,\sigma} +
h.c. ) - g\sum_{i,\sigma} (n_{i,\sigma}-\frac{1}{2})(a_i +a^{\dagger}_i) +
\nonumber\\
&+& \omega_0 \sum_i a^{\dagger}_i a_i,
\end{eqnarray}
where $c_{i,\sigma}$ ($c^{\dagger}_{i,\sigma}$) and $a_i$ ($a^{\dagger}_i$) are, respectively,
 destruction (creation) operators for
fermions and for local vibrations of frequency $\omega_0$
on site $i$, $n_{i,\sigma}=c^{\dagger}_{i,\sigma}c_{i,\sigma}$
the electron density per spin, $t$ is
the hopping amplitude, $g$ is an electron phonon coupling.
In the half-filled case we always fix the chemical potential to the particle-hole
symmetric value, which fixes the density per spin to $n = 1/2$.
In the spinless case there is no sum on $\sigma$.
We choose as parameter of the model the e-ph coupling constant
$\lambda = 2g^2/\omega_0 D$ where $D$ is the half-bandwidth of our infinite-coordination
Bethe lattice, and the adiabatic ratio $\ad = \omega_0/D$.
In the single electron case the spin index is unessential, moreover the mean density is
zero and the el-ph interaction in Eq. (\ref{eq:themodel}) is replaced by
$- g\sum_{i} n_{i}(a_i +a^{\dagger}_i)$.
The partition function (\ref{Zpart}) is now defined as
\beq
\label{Zholstein}
Z = \int \mathcal{D} x(\tau) \int \mathcal{D} \psi^{\dagger} \mathcal{D} \psi(\tau)  \exp(-S)
\eeq
where using units where the spring constant $K=M\omega_0^2=1$ 
$x(\tau)=\sqrt{\omega_0/2}(a(\tau)+a^\dagger(\tau))$.

The single impurity action $S$ associated to the lattice Hamiltonian (\ref{eq:themodel}) in
an infinite coordination lattice reads this case:
\begin{eqnarray}
\label{Sel}
S &=& -\int_0^\beta d\tau \int_0^\beta d\tau' \sum_{\sigma}
\psi_{\sigma}^\dagger(\tau) \mathcal{G}^{-1}_0(\tau-\tau') \psi_{\sigma}(\tau') +
\\
\label{Sph}
 &+&\frac{1}{2} \int_0^\beta d \tau \left( \frac{\dot{x}^2(\tau)}{\omega_0^2} +
 x^2(\tau) \right) +
\\
\label{Selph}
&+&\sqrt{\lambda} \int_0^\beta d\tau x(\tau) \left(n(\tau) - 1 \right).
\end{eqnarray}
where $n(\tau)=\sum_{\sigma} \psi_{\sigma}^\dagger(\tau) \psi_{\sigma}(\tau)$.

\subsection{DMFT-QMC method}
\label{sec:DMFT-QMC}
An efficient method to solve Eqs. (\ref{Sel},\ref{Sph},\ref{Selph}) at finite temperature is 
the QMC method (in the Blankenbecler-Scalapino-Sugar
approach\cite{BSS}. This method works well for not too low temperatures and naturally yields
electronic and bosonic correlation
functions, as well as the probability distributions associated to the phonon
fields. The method is not affected by the negative
sign-problem, and its main limitation comes in the adiabatic regime
($\gamma \ll 1$) where the phonon becomes heavy making 
more difficult to sample correctly the available phase space. 
In the BSS scheme the fermions
are integrated out , and the phonons coordinates $x(\tau)$ are
discretized in $L$ imaginary-time slices of width $\Delta \tau = \beta/L$ and
then sampled by QMC.
$L$ has to be chosen large enough to reduce as much as possible
$\Delta\tau$, which  controls the the Trotter discretization error.
To keep $\Delta \tau$
less than $1/8$ we used  $32$ slices except for the 
lowest temperature ($\beta=8$) for which we have used $L=64$,

\subsection{DMFT-ED method}
\label{sec:DMFT-ED}

The Anderson Impurity Model for the Holstein model reads
\begin{eqnarray}
\label{eq:Anderson_Holstein}
H_{AIM} &=& = -\sum_{k,\sigma} V_k (c^{\dagger}_{k,\sigma} f_{\sigma} + h.c)+
\sum_{k,\sigma} E_k c^{\dagger}_{k,\sigma} c_{k,\sigma} -\nonumber\\
&-&g \sum_\sigma \left ( f^\dagger_\sigma f_\sigma - \frac{1}{2}\right) (a +a^{\dagger}) + \omega_0 a^{\dagger}  a.
\end{eqnarray}
We solve (\ref{eq:Anderson_Holstein}) by means of ED by truncating the sums
in the first two terms of Eq. (\ref{eq:Anderson_Holstein}) to a small number
of terms $N_b$, so that the Hilbert space is small enough to use, e.g., the
Lanczos algorithm to compute the $T=0$ Green's function.
For the case of phonon degrees of freedom we consider here, also the infinite
phonon space has to be truncated allowing for a maximum number of excited
phonons $N_{ph}$. In all the calculations presented here the convergence of
both truncations have been checked. The value of $N_{ph}$ has to
be chosen with special care in the adiabatic regime and in strong coupling,
where phonon excitations are energetically convenient.
As far as the discretization of the bath is concerned, the convergence of
thermodynamic averages and Matsubara frequency properties is exponentially
fast and $N_b \sim 8-9$ is enough to obtain converged results.
The method also offers the advantage of a direct evaluation of real-frequency
spectral properties such as the electron and phonon spectral functions.
The main limitation is that these quantities reflect the discrete nature of
our system. In practice,
the spectra are formed by collections of $\delta$-functions, which 
limits our frequency resolution, and makes the method better
suited to gain knowledge on the main features of the spectra, rather than
on the fine details.

\subsection{Quantities of interest}
\label{sec:quantities}
We mainly characterized electronic and phononic properties 
by considering respectively the electron density of states
$\rho(\omega) = -\frac{1}{\pi} G(\omega)$ and the 
phonon probability distribution function (PDF) 
$P(x) = \left \langle\phi_0 |x\rangle \langle x| \phi_0\right \rangle$,
where $|x\rangle$ is a phonon coordinate eigenstate, and $|\phi_0\rangle$ 
is the ground state vector.
At finite electron density a lattice polarization reflects in 
the presence of two peaks in $P(x)$, 
corresponding to opposite polarization of
occupied and unoccupied sites (bimodal behavior) \cite{Millis-adiab}.
In the single electron case the polarization of a lattice site where one
electron sits is a marker of a polaronic crossover. 
Also in this case a bimodal behavior can be observed in the adiabatic regime 
but generally speaking we have a definite polarization when phonon fluctuations are less than
the average polarization due to the presence of the electron.
In this way a {\it qualitative} difference is identified between the 
polarized and unpolarized regimes, which allows for an unambiguous way
to draw a crossover line, as opposed to estimates based on smoothly
varying functions as average lattice fluctuations
or electron kinetic energy.

A Metal to Insulator Transition (MIT) can be probed by the low energy behavior of
$\rho$. This is a consequence of the infinite dimensional limit
where the self-energy is momentum-independent. Thus the vanishing of the 
low-energy quasi-particle spectral weight
coincides with the divergence of the effective mass, which determines the MIT.
 
\section{Single electron in Holstein and Holstein t-J models}
\label{sec:HolsteinAndHtJ}

\subsection{$T=0$ continued fraction for a single polaron}
\label{sec:singlepolaron}
The single electron case in both Holstein, $t-J$ and $t-J$-Holstein case can be
solved semi-analytically taking advantage of peculiar features of the
zero density case. We briefly describe here the formalism at $T=0$. Generalization to
thermalized lattice can be found in \cite{depolarone}. 
We use here the AIM
formalism of eq. (\ref{eq:Anderson_Holstein}).
For a single electron the Green's function 
is purely retarded, then the retarded 
impurity propagator can be 
defined as
\begin{equation} 
\label{Gfreq}
G(\omega)=\langle0|f\frac{1}{\omega+i\delta-H}f^\dagger|0\rangle
\end{equation}
which has the correct prescription $\delta>0$ for convergence of the time
integrals. The vacuum energy is defined here to be zero.
To proceed further one needs to introduce the
generalized matrix elements
\begin{equation}
\label{matrix-elements}
G_{n,m}=\langle0|\frac{a^n}{\sqrt{n!}}f \frac{1}{\omega+i\delta-H}f^\dagger
\frac{(a^\dagger)^m}{\sqrt{m!}}|0\rangle
\end{equation}
so that the element $G_{0,0}$ will be the Green function.

Let us separate the impurity Hamiltonian of eq. (\ref{eq:Anderson_Holstein}) into  $H_0$ and
$H_I$,  $H_I$
is the local interaction term and $H_0$ the remainder. 
A useful operator identity for the resolvent is
\begin{equation} \label{Risolv}
\frac{1}{z-H}=\frac{1}{z-H_0}+
                   \frac{1}{z-H_0} H_I \frac{1}{z-H}
\end{equation}
The diagonal matrix element of this operator on the impurity
zero phonon state
$f^\dagger |0\rangle$ is the Green's function of eq.
(\ref{Gfreq}). 

In the subspace of zero electron $p$-phonon states
$|0,p\rangle=(a^\dagger)^p/\sqrt{p!}|0\rangle$
one can write
\begin{equation}
H_I=\sum_p f^\dagger |0,p\rangle\langle 0,p|f (a+a^\dagger)
\end{equation}
leading to the recursion formula for the $G_{n,m}$'s
\begin{equation} \label{Eqric}
G_{n,m}=G_{0n} \delta_{n,m} -g\sum_p G_{0n} X_{n,p} G_{p,m}
\end{equation}
where  $G_{0n}=G_0(\omega-n\omega_0)$ is the diagonal element of the
free resolvent and $X_{n,p}$ are the phonon displacement matrix
elements $X_{n,p}= \sqrt{p+1} \delta_{n,p+1}+ \sqrt{p} \delta_{n,p-1}$.

One immediately recognizes that, due to the particular form of the matrix ${\bf X}$,
${\bf G}^{-1}$ is a tridiagonal matrix, so
that the solution of the problem is reduced to the inversion of a matrix
in arbitrary dimensions.
Following the lines given in Ref.\cite{Viswanath-Muller}
one can
express the diagonal element of the ${\bf G}$ matrix in terms of the
diagonal and non-diagonal elements of ${\bf G}^{-1}$.
The local propagator (the $0,0$ element of ${\bf G}$) is obtained in
terms of a Continued Fraction Expansion (CFE), as a functional
of the ''bare'' propagator $G_0$:
\begin{equation}
\label{CF}
G(\omega)={1 \over\displaystyle G_0^{-1}(\omega)-
{\strut g^2 \over\displaystyle G_0^{-1}(\omega-\omega_0)-
{\strut 2g^2 \over\displaystyle G_0^{-1}(\omega-2\omega_0)-
{\strut 3g^2 \over\displaystyle G_0^{-1}(\omega-3\omega_0)-...}}}}
\end{equation}
Due to the impurity analogy,
this is also
the local propagator of the original lattice problem, provided
that self-consistency condition (eq. (\ref{selfcons2}) for real frequencies)
is fulfilled.  As a consequence the lattice 
self-energy $\Sigma$ is immediately obtained by $G=1/(G^{-1}_0-\Sigma)$.
An example of the spectral function that can be obtained with this formalism 
is presented also in this volume (see
\cite{FehskeBronoldAlvermannVarenna} fig. 4).

\subsection{Holstein $t-J$ model in infinite dimensions.} 
\label{sec:HtJ}
%The case of $t-J$ model with
%additional Holstein-like electron-phonon interaction is a little bit more
%involved.
%Let us consider the Holstein-$t$-$J$ model defined by the
%Hamiltonian:\cite{martinez,ramsak}
%\begin{eqnarray}
%H&=&-\frac{t}{2}\sum_{\langle ij \rangle \sigma}
%\left(\tilde{c}^\dagger_{i \sigma}\tilde{c}_{j \sigma} + {\rm h.c.}\right)
%+\frac{J}{2}\sum_{\langle ij \rangle}
%\left[\left(S_i^z S_j^z -\frac{n_in_j}{4}\right)
%+\frac{1}{2}\left(S_i^+ S_j^- + S_i^- S_j^+ \right)\right]\nonumber\\
%&&+g \sum_i \left[ n_i - \langle n_i \rangle \right] (b_i+b_i^\dagger)
%+\omega_0\sum_i b_i^\dagger b_i,
%\label{hamil1}
%\end{eqnarray}
%\endwide
%where $\tilde{c}^\dagger_{i \sigma}$ are the electron operators
%in the presence of infinite strongly correlations that prevents
%double occupancy [$\tilde{c}^\dagger_{i \sigma} 
%= c^\dagger_{i \sigma}(1-n_{-\sigma})$],
%$b^\dagger_i$ the phonon operators and
%$S^{z,+,-}_i$ respectively the $z$ component
%and the raising and lowering spin operators.

To derive the Holstein-$t-J$ Hamiltonian for a single hole we sketch 
the treatment of Ref. \cite{HtJ}.
First the $t-J$ Hamiltonian is transformed by
a canonical transformation into
a ferromagnetic one.
Then we introduce fermionic $h$ {\it hole} 
and bosonic $b$ {\it spin-defect on the antiferromagnetic
ground state} 
operators.
As a further step an Holstein-Primakoff transformations is performed which
introduces spin waves \cite{martinez}, then we adopt the linear spin wave scheme and we use
explicitly the infinite coordination limit to get:
\begin{eqnarray}
H&=&\frac{t}{2\sqrt{z}}\sum_{\langle ij \rangle \sigma}
\left(h_j^\dagger h_i a_j + {\rm h.c.}\right)
-g \sum_i \left[h_i^\dagger h_i  - \langle h_i^\dagger h_i \rangle \right]
(a_i+a_i^\dagger)
+\omega_0\sum_i a_i^\dagger a_i\nonumber\\
&&+\frac{J}{4z}\sum_{\langle ij \rangle}
\left[b_i^\dagger b_i+b_j^\dagger b_j\right]
+\frac{J}{2} \sum_i h_i^\dagger h_i.
\label{hamilhatjz}
\end{eqnarray}
The first term of Eq. (\ref{hamilhatjz}) describe the kinetic hopping of one
hole on the antiferromagnetic background, which is accompanied by the creation
(destruction) of a spin defect which breaks (restores) $2z$ magnetic bonds with
individual energy $J/4z$. In addition we have the usual local e-ph
interaction which couples {\rm the hole density} to the local phonon. The last
term in Eq. (\ref{hamilhatjz}) can be absorbed in the definition of the hole
chemical potential which, for the single hole case here considered, has to be
set at the bottom of the hole band. 
The single-hole Green's function is given
by the resolvent of Eq. (\ref{Gfreq}) in which the spinless fermions $f$ 
are replaced now by the hole operators $h$.
Following the same path of the previous calculation 
and using the results of Ref. \cite{strack} for the $t-J$ model,
we can write the local hole propagator
$G(\omega)$ as function of the sum of a hopping
and a phonon contribution to the self-energy\cite{HtJ}:
\begin{equation}
\label{eqG}
G(\omega) = \frac{1}{\omega - \Sigma_{\rm hop}(\omega)-\Sigma_{\rm el-ph}(\omega)},
\end{equation}
where the hopping contribution describes the dynamics
of the hole through the antiferromagnetic background,
\begin{equation}
\Sigma_{\rm hop}(\omega)  = \frac{t^2}{4} G(\omega-J/2),
\label{sigmat}
\end{equation}
while the e-ph self-energy takes into account the on site multiple 
scattering with phonons and it is formally the same as in Eq. (\ref{CF})
after identifying
\begin{equation}
\label{Gt}
G^{-1}_0(\omega)=\omega-\frac{t^2}{4} G(\omega-J/2).
\label{gt}
\end{equation}
Both $\Sigma_{\rm hop}(\omega)$
and $\Sigma_{\rm el-ph}(\omega)$ are expressed as
functionals of the {\em total} Green's function $G(\omega)$ leading
to a self-consistent interplay between the spin and the e-ph
interaction.
Eqs. (\ref{eqG}), 
(\ref{sigmat}) and (\ref{gt}) represent a closed self-consistent system
which we can be numerically solved by iterations to obtain the explicit
{\em exact} expression of the local 
Green's function $G(\omega)$ \cite{HtJ}.
The formal scheme looks quite similar to the single electron solution
of the Holstein model \cite{depolarone}.
However, due to the antiferromagnetic background,
the physical interpretation is quite different.

Due to the orthogonality of the initial and final
antiferromagnetic background, the non-local component
of the Green's function in the Holstein-$t-J$ model,
even  for $J=0$,
is strictly zero $G_{ij}(\omega)=G(\omega)\delta_{ij}$,\cite{strack} whereas
for the pure Holstein model $G_{i \neq j}(\omega)$ is finite and
provides informations about the non-local
dynamics: 
$G({\bf k},\omega)=1/[\omega -\epsilon_{\bf k}-\Sigma(\omega)]$.

In addition, the magnetic ordering has important consequences
also on the local Green's function $G_{ii}(\omega)$.
In the pure Holstein model for instance $G_{ii}(\omega)$ takes
into account any generic dynamics which occurs back and forth a given site 
whereas
in the Holstein-$t$-$J$ model the electron must follow
a retraceable path in order to restore the antiferromagnetic
background.\cite{strack}
A Bethe-like dynamics is thus enforced by the magnetic ordering regardless
the actual real space lattice.
The object made up by the
hole plus the local modification of the spin configuration due
to the presence of the hole is the ``spin polaron''.

\subsection{Results}
\label{sec:resultsHtJ}
\begin{figure}[htbp]
\begin{center}
\includegraphics[scale=0.33,angle=270]{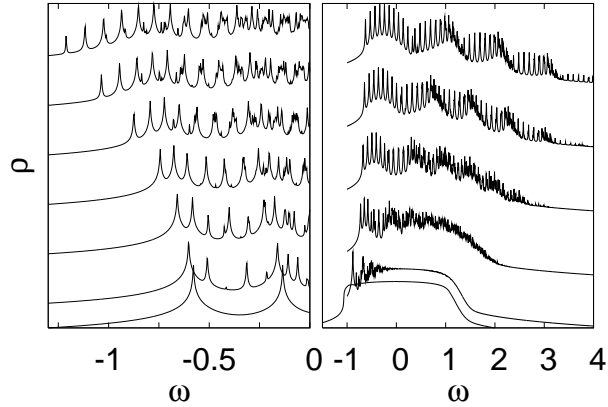}
\end{center}
\caption{Hole spectral function in the $t-J$-Holstein model for $\gamma=0.1$
as function of $J/t$ and $\lambda$.
Left panel: $J/t=0.4$ and (from bottom to top) 
%$\lambda=0.0,0.1,0.3,0.5,0.7,0.9,1.1$.
$\lambda=0.0,0.05,0.15,0.25,0.35,0.45,0.55$.
Right panel: $\lambda=0.25$ and (from bottom to top)
$J/t=0.0,0.1,0.5,1.0,1.5,2.0$.}
\label{fig:DOS_HtJ}
\end{figure}

The local physical properties of one hole in the infinite dimension
Holstein-$t-J$ model have been extensively investigated in Ref. \cite{HtJ}.
In Fig.~\ref{fig:DOS_HtJ} we report the evolution of the local spectral
function $\rho(\omega)=-(1/\pi)\mbox{Im}G(\omega)$ as
function of the spin-exchange $J/t$ and of the e-ph interaction
$\lambda$ for $\omega_0/t=0.1$.
In the limits $J=0$ and $\lambda=0$ we recover respectively
the results of Refs. \cite{depolarone} and \cite{strack}.
They are qualitatively different: for $J/t=0$ and finite $\lambda$
the spectral density is the same as the pure Holstein model (see e.g. fig. 5 
\cite{FehskeBronoldAlvermannVarenna} of the present volume),
whereas for $\lambda=0$ and finite $J/t$ the spectrum
is described by magnetic peaks
spaced as $(J/t)^{2/3}$ (for small $J/t$).
Switching on at the same time both the magnetic interaction and the
e-ph interaction gives rise to
the interesting interplay between these degrees of freedom.
This can be shown for instance in the evolution of the spectral function
as function of $\lambda$: increasing the e-ph interaction not only gives
rise to additional phonon peaks which superimpose on the magnetic one, but it 
also enhances the nature of the magnetic polaron, from a large one
(corresponding to $(J/t)^{2/3}$ spaced peaks) to a more localized
one (corresponding to $(J/t)$ spaced peaks).
A similar behavior appears as function
of $J/t$: here for large $J/t$ the gross structure of the
spectral function is described by equally ($J$) spaced
magnetic structures with fine features determined by the e-ph interaction.
Note that this latter is represented by phononic peaks spaced by $\omega_0$ as in the
antiadiabatic limit, although $\gamma$ here is just $\gamma=0.1$, since the
 antiadiabatic regime is intrinsically enforced by the reduction of the kinetic energy
 due to the magnetic polaron trapping.
In any case within DMFT we loose hole band dispersion. As a consequence we were
not able to reproduce even qualitatively the interchange of magnetic and
polaronic peaks observed by increasing el-ph coupling at finite dimensionality
\cite{MishchenkoVarenna}. However within our localized solution we recover
many of the qualitative features of both magnetic and 
lattice polaron crossovers \cite{HtJ}.
 
\begin{figure}[htbp]
\begin{center}
\includegraphics[scale=0.33,angle=270]{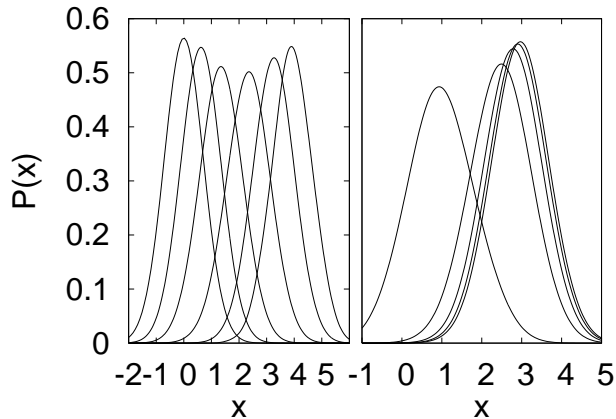}
\end{center}
\caption{Phonon PDF in the $t-J$-Holstein model for $\omega_0/t=0.5$
Left panel:
$J/t=0.4$ and (from left to right) 
%$\lambda=0.0,0.1, 0.5, 1.0, 1.5, 2.0, 2.5$.
$\lambda=0.0,0.05, 0.25, 0.5, 0.75, 1.0, 1.25$.
Right: $\lambda=0.5$ from left to right $J/t=2.0, 1.0, 0.4 0.05$.}
\label{fig:HtJ-px}
\end{figure}

 Another interesting quantity which points out the interplay between magnetic
and e-ph interaction is the  PDF $P(x)$.
Notice due to localization of DMFT solution $P(x)$ gives the lattice distortion
associated to the localization center.
In the left panel of Fig. \ref{fig:HtJ-px}
we show the $P(x)$ for $J/t=0.4$ as function of $\lambda$.
Here the lattice polaron formation is characterized by 
the value $\lambda_{pol}$ at which 
broadening of the
PDF is maximum. 
For larger $\lambda$ the PDF recovers a gaussian form due to
the lattice fluctuations around the new minima at finite distortion. The
magnetic polaron formation is also pointed out by the analysis of the $P(x)$
(right panel of Fig. \ref{fig:HtJ-px}): the magnetic trapping favors the
lattice one, further  reducing the anomalous lattice
fluctuations towards the gaussian ones.

\section{Half-filled Holstein model: spinless vs spinful cases at $T=0$}
\label{sec:HHFED}

\subsection{Adiabatic regime}
\label{sec:HHFEDad}
We briefly describe, as a starting point of the Born-Oppenheimer (BO) procedure, the
adiabatic limit in which $\ad \rightarrow 0$ keeping
$\lambda$ fixed. This limit has been thoroughly studied in  Ref.
\cite{Millis-adiab} which we here  briefly resume. 
When $\omega_0\rightarrow0$ and $\lambda$ finite 
the kinetic term (\ref{Sph})
forces the phonon path $x(\tau)$ to be $\tau$-independent, $x(\tau)\equiv x$. Phonons becomes 
classical and the interaction term reads  $-\sqrt{\lambda} x \int_0^\beta
d\tau (n(\tau)-1)$. 
The gaussian integrals in
(\ref{Zholstein}) with the action given by (\ref{Sel},\ref{Sph},\ref{Selph}) can be computed analytically leading
to 
\beq
Z = \int d x \exp(-\beta V(x))
\eeq
where the adiabatic potential $V(x)$ is \cite{NoteKF2,BrandtMielsch,ChungFreericks}
\begin{equation}
V(x)=\frac{1}{2} x^2-\frac{\sqrt{\lambda}}{2} |x|-\frac{s}{\beta}\sum_n
\log \left ( \frac{G^{-1}_0(i\omega_n)+\sqrt{\lambda} x}{i\omega_n+\sqrt{\lambda} x}\right )
\label{eq:E-adiab}
\end{equation}
where $s$ is spin degeneracy.
In formula Eq. (\ref{eq:E-adiab}) we have found useful to separate the
contribution in absence of hybridization (
first line of Eq. (\ref{eq:E-adiab})) 
from a remainder (last line). 
Through the adiabatic potential we compute the phonon PDF as
\beq
\label{defPXadiab}
P(x) = \frac{\exp(-\beta V(x))}{Z}.
\eeq

Taking advantage of the Gaussian nature of the fermions we get for the local propagator
\beq
\label{Gadiab}
G(\omega) = \int d x P(x) \frac{1}{G^{-1}_0(\omega)-\sqrt{\lambda} x}.
\eeq
which defines the self consistency condition through Eq. (\ref{selfcons2}).

The self-consistency condition (\ref{selfcons2}) together with
Eqs. (\ref{Gadiab},\ref{defPXadiab},\ref{eq:E-adiab}), completely solves the problem.
Notice also  the correspondence  between the spinless and and the spinful case
upon rescaling $\lambda$ to $\lambda/2$ in the latter case \cite{Millis-adiab}.
For the Bethe lattice (at zero temperature) 
it can be shown that the potential
(\ref{eq:E-adiab}) becomes double welled above a critical value 
$\lambda_{pol}=3\pi/(8 s)$. A MIT occurs at a larger coupling
$\lambda_{MIT} = 1.328/s$\cite{Millis-adiab}.

The BO procedure goes on by quantizing the adiabatic potential after
adding the phonon kinetic energy contribution. Introducing the scaled
variable $u=gx/\sqrt{s}$ the BO phononic Hamiltonian reads
\beq
\label{eq:BO}
H_{BO}=-\frac{\ad}{2s}\frac{d^2}{d u^2}+s V(u).
\eeq
and $V(u)$ is given by Eq. (\ref{eq:E-adiab}).
Notice that the spinful BO Hamiltonian
maps onto twice the spinless one upon
rescaling 
\begin{eqnarray}
\label{scaling}
\lambda/2 &\rightarrow& \lambda \nonumber \\
2\ad &\rightarrow& \ad.
\end{eqnarray}

%Whether the gradient term represents the most relevant contribution from
%quantum fluctuation of phonons can be questionable. A more general non-local
%contribution to the adiabatic potential arises in the effective phonon action
%as it is discussed in Ref. \cite{pata}. As it is discussed there this non local
%part may affects the determination of the phonon properties. However we decide
%to pursuit the way of simplicity and discuss the BO approximation in view of
%comparison with ED data.

While phonon  properties are immediately obtained at this stage from the
solution of the one-dimensional anharmonic system of Hamiltonian  Eq.
(\ref{eq:BO}), electronic properties must account non-trivially
for the tunneling of phonon coordinates.

The simplest way to describe electrons coupled to a tunneling
system is to map it onto a two level system. In our model this can be
accomplished by changing the basis (operators $a$) from that of the
harmonic oscillator to the that defined by the solution of (\ref{eq:BO}).
Then projecting
out all the states but the first two ($\vert +\rangle,\vert -\rangle$) we get the following two
state projected model (TSPM):
\begin{eqnarray}
\label{eq:TSPM-definition}
H&=&-\frac{2}{s}\sum_\sigma\e\left(f_\sigma^+f_\sigma-\frac{1}{2}\right)\sz -\D\sx
+\sum_{k,\sigma}E_k c^\dagger_{k,\sigma}c_{k,\sigma}+\nonumber\\
&+&\sum_{k,\sigma} V_k \left( f_\sigma^\dagger c_{k,\sigma} + c^\dagger_{k,\sigma} f_\sigma\right),
\end{eqnarray}
where $\sigma_x$ and $\sigma_z$ are two Pauli matrices in the space spanned by $\vert + \rangle$
and $\vert - \rangle$ and the quantities $\epsilon$ and $\Delta$ are given by
\begin{eqnarray}
\label{eq:TSPM-parameters}
 \epsilon &=& g\frac{s}{2} \langle+\vert a+a^\dagger\vert -\rangle\\
 \Delta &=& \frac{\omega_0}{2}(\langle +\vert a^\dagger a\vert +\langle - \rangle -\vert a^\dagger a\vert -\rangle)
\end{eqnarray}
The latter quantity $\Delta$ is the tunneling frequency
between the two phononic states.
A similar model has been introduced in Ref. \cite{pata} to study the strong
coupling limit of the Holstein model.
The  TSPM reproduces exactly the DMFT of the Holstein model in two limits: 
weak coupling and adiabatic limit.  In the
former case the projection of the phonon space has no relevance therefore the
TSPM reproduces the perturbation expansion developed (in the limit of infinite
bandwidth) in Ref. \cite{Engelsberg}. The adiabatic limit is
instead recovered as $\Delta\rightarrow 0$. No phonon tunneling occurs and the
model can be solved exactly by CPA recovering the solution of Ref.
\cite{Millis-adiab}.

To analytically span from the strong ($V_\kvec \rightarrow 0$) to the  weak
($g\rightarrow 0$) coupling regimes of the equivalent impurity Hamiltonian it
is useful to devise an Iterated Born-Oppenheimer Coherent Potential Approximation (IBOCPA) scheme.
Starting from the Green's function for $V_k=0$
(\ref{eq:GatBOCPAspinless}) for the spinless case and (\ref{eq:GBOCPAspinful}) for the spinful
one, we notice that in both cases $G_a(\o)$ can be written ad sum of two contribution
$G_a(\o)=(1/2)(G_{a,+}(\o)+G_{a,-}(\o))$ where
($\pm$) label a phonon state \cite{PCarta}.
Then we write the propagator in presence of hybridization as
\beq
\label{eq:GBOCPA}
G(\o)=\frac{1}{2}\sum_{\s=\pm}\frac{1}{G^{-1}_{a,\s}(\o)-\frac{D^2}{4}G(\o)}
\eeq
where the hopping of the electron from the impurity to the
bath is described by the cavity-field correlator $D^2/4
G(\o)$ (\ref{selfcons2}).
Iteration proceeds substituting  the propagator (\ref{eq:GBOCPA})  back in 
Eq.  \ref{eq:E-adiab} giving a new BO potential and  new
TSMP parameters (eq. (\ref{eq:TSPM-parameters}))  and finally a new Green's
function. Iteration continues after convergence is reached \cite{NotaIBOCPA}.

In (\ref{eq:GBOCPA}) $G^{-1}_{a,\s}(\o)$ are propagators obtained from the solution of
the atomic ($V_k=0$) limit of the TSPM in both spinless and spinful cases.
The spinless atomic Green's function can be easily found to be
\beq
\label{eq:GatBOCPAspinless}
G_a(\o) = \frac{1}{2}\sum_{\s=\pm}
\left( \frac{\e^2}{\l^2}\frac{1}{\o+2\l\s} +
\frac{\D^2}{\l^2}\frac{1}{\o}\right)
\eeq
$G_a(\omega)$ has a pole at $\omega=0$ induced by phonon tunneling, whose weight vanishes  as $\Delta\rightarrow 0$, accompanied by
two resonances at $\pm 2\l$.
The zero energy peak is due to transitions in which both charge and phonon "spin"
change while the side peaks arise from charge transfer in a frozen phonon "spin".
In this sense the side peaks are adiabatic features which survive when phonon tunneling
$\Delta\rightarrow 0$.
The spinful atomic  Green's function is
\beq
\label{eq:GBOCPAspinful}
G_a(\o)=\frac{1}{2}\sum_{\s=\pm}\frac{1}{2\l}\left(
\frac{\l-\D}{\o+\s(\l+\D)}+\frac{\l+\D}{\o+\s(\l-\D)}\right)
\eeq
The most striking difference with the spinless case (\ref{eq:GatBOCPAspinless}) is the
absence of the zero frequency pole. In the spinful case the tunneling of the
phonon $\Delta$ is always associated to a finite energy transition, and it only splits the
finite frequency poles associated to the transition from singly  to
the empty or doubly occupied ones.

IBOCPA assumes that the tunneling states of the phonon in the adiabatic
potential remain unaltered during an hybridization event. 
As in standard CPA a band is associated to 
each local level, but  our  IBOCPA gives a Fermi liquid solution in the spinless case
for every value of the coupling, as opposed to the case of the Hubbard model. 
The low-energy band arising from the zero
energy pole in the zero hybridization limit is indeed {\it coherent}. This can
be easily realized by analysis of  the self-energy. When $\omega\rightarrow 0$
the spinless propagator defined by  (\ref{eq:GatBOCPAspinless})
and (\ref{eq:GBOCPA}) is dominated by the zero-energy pole of $G_a$ 
(\ref{eq:GatBOCPAspinless}) and consequently the self-energy obtained through
(\ref{eq:GBOCPA}) is purely real.

Conversely in the spinful case a MIT due to local pair formation occurs 
in the CPA approximation at a critical value of $\lambda$.
Finally we emphasize that we recover the adiabatic
solution of Ref. \cite{Millis-adiab} as $\ad \rightarrow 0$ for finite
$\lambda$ as $\Delta\rightarrow 0$. In this case the IBOCPA is exact and gives
the Green's function of Ref. \cite{Millis-adiab}. However the IBOCPA procedure
is certainly affected by serious problems approaching the MIT in the spinful
case. A more careful treatment of the low-energy part of the Green's function
has been performed in this case in Refs. \cite{pata,Bulla,HewsonVarenna}, where it
has been observed a MIT scenario similar to the half-filled the Hubbard model,
i.e., a quasi-particle peak that shrinks to zero width approaching a critical
value of $\lambda$. In
the spinless case instead a resonance is present at zero energy
within a CPA approach. It is not associated to a Kondo effect but rather to
phonon tunneling which drives charge fluctuations. On the other hand a Kondo
like behavior can be ascribed to the bipolaron or pair formation. For a
discussion of the limitation of the IBOCPA approach  see also Ref.
\cite{PCarta}.

\subsection{Antiadiabatic regime}
\label{sec:HHFEDantiad}
While in adiabatic limit the phonon displacement becomes a classical variable,
and we are left with an electronic model which depends parametrically on it, in
the opposite limit ($\ad >> 1$) the roles are exchanged, and we have a
parametrically fixed electronic charge on a given site. In this regime the most
reasonable starting point is  the Lang-Firsov (LF) canonical
transformation\cite{Lang-Firsov,JuliusVarenna} $S=\exp (T)$.
The generator of the transformation reads
\beq
\label{eq:LangFirsov}
T = -\alpha \sum_\sigma (f^\dagger_\sigma
f_\sigma-\frac{1}{2})(a^\dagger-a),
\eeq
introducing the parameter $\alpha=g/\omega_0$ which is the relevant e-ph
coupling parameter in the anti-adiabatic regime \cite{storia,depolarone}.

The canonical transformation diagonalizes the impurity Hamiltonian in the absence
of hybridization by eliminating the e-ph interaction part.
In the spinful case the phonon energy term of (\ref{eq:Anderson_Holstein})
gives rise to the well known bipolaronic instantaneous attraction
\cite{JuliusVarenna}.
The hybridization
term of (\ref{eq:Anderson_Holstein}) is modified 
by acquiring an exponential term in the phonon
coordinates leading to
\begin{eqnarray}
\label{eq:AndersonLF}
e^T H e^{-T} = -\sum_{k,\sigma} e^{\alpha (a^\dagger-a)}
V_k (c^{\dagger}_{k,\sigma} f_{\sigma} +
h.c.) +
\sum_{k,\sigma} E_k c^{\dagger}_{k,\sigma} c_{k,\sigma}-\nonumber \\
-2\frac{g^2}{\omega_0} (s-1) n_\uparrow n_\downarrow -
\frac{g^2}{\omega_0} \sum_\sigma (\frac{1}{2}- n_\sigma)
 + \omega_0 a^{\dagger}  a,
\end{eqnarray}
where $n_{\sigma}=f^{\dagger}_{\sigma} f_{\sigma}$.
Notice that in the anti-adiabatic limit  $\ad \rightarrow \infty$, if $\lambda$
is kept constant $\alpha$ vanishes. In this case spinless electrons are not
renormalized, while spinful electrons are described by an attractive Hubbard
model with $|U|/D=\lambda$.

If we want to proceed with analytical methods, the hybridization term must be 
treated in an approximate way. Assuming that in the anti-adiabatic limit the
impurity density is constant during the fast motion of the phonon, we average
out the phonon term on the displaced phonon ground state. This is the so-called
Holstein Lang-Firsov Approximation (HLFA), which has not to be confused with
the exact canonical transformation (\ref{eq:LangFirsov}). HLFA gives
rise to the exponential renormalization of the hybridization constants where
each $V_k$ is replaced by $V_k\exp(-\alpha^2/2)$.
Such a replacement implies the well known exponential renormalization of the bandwidth
$D\exp(-\alpha^2)$.

To get the {\it electron} Green's function $G(\omega)$, 
the explicit action of LF transformation
have to be taken into account into both creation and destruction operators
appearing in the definition of the Green function. 
Following Refs. \cite{Ranninger_spectral} and
\cite{Alexandrov-Ranninger}, we obtain in both spinless and spinful cases
\beq
\label{eq:GreenLF}
G(\omega)=e^{-\alpha^2}G_p(\omega)
+\frac{1}{2}\sum_{n\ne0}e^{-\alpha^2} \frac{\alpha^{2|n|}}{|n|!} G_p(\omega-n\omega_0).
\eeq
where $G_p(\omega)$  is the Green's function of an impurity (with a  negative $U$
interaction in the spinful case) with an exponentially reduced
hybridization to a bath of conduction electrons. The self-consistency condition can
be written explicitly in the spinless case due to the lack of interaction terms on
the impurity:
\beq
\label{eq:Gp_spinless}
G_p(\omega) = \frac{1}{\omega-e^{-\alpha^2}\frac{t^2}{4}G(\omega)}.
\eeq
where $G(\o)$ is the local Green's function of the lattice.
In the spinful case a Lang-Firsov Coherent Potential Approximation (LFCPA)
can be devised for the resulting HLFA
attractive Hubbard model with  $U=-2g^2/\omega_0$ giving
\beq
\label{eq:Gp_spinful}
G_p(\omega) = \frac{1}{2} \left (
\frac{1}{\omega-e^{-\alpha^2}\frac{t^2}{4}G(\omega)-U/2}+ \right.
 \left. \frac{1}{\omega-e^{-\alpha^2}\frac{t^2}{4}G(\omega)+U/2}
\right ).
\eeq

%A comparison between the zero hybridization (atomic $D=0$) DOS and
%results from HLFA and LFCPA results is depicted in figs.
%\ref{fig:LFspinless}, \ref{fig:LFspinful}.

%\begin{figure}[htbp]
%\begin{center}
%\includegraphics[scale=0.33,angle=270]{fig_LF_spinless.eps}
%\end{center}
%\caption{(color online) The spectral function in the spinless case in the zero hybridization case
%(bold line)  and in the HLFA approximation (thin line). The upper panel
%refers to a typical weak-coupling situation ($\lambda=2.7,\ad=1.2$)
%while the lower panel shows results for intermediate
%coupling ($\lambda=5.4,\ad=1.2$).}

%\label{fig:LFspinless}
%\end{figure}

%\begin{figure}[htbp]
%\begin{center}
%\includegraphics[scale=0.33,angle=270]{fig_LF_spinful.eps}
%\end{center}
%\caption{(color online) The spectral function of the spinful case in the zero
%hybridization case (bold line)  and in the LFCPA approximation (thin
%line). The upper panel is dedicated to weak coupling ($\lambda=2.7,\ad=1.2$)
%and the  lower panel to 
%intermediate coupling
%($\lambda=5.4,\ad=1.2$).}
%\label{fig:LFspinful}
%\end{figure}
Notice that the theory developed here for the Holstein impurity model differs
from that developed directly in the lattice
model\cite{Ranninger_spectral}. In that case an equation identical to
(\ref{eq:GreenLF}) is recovered for a band of free electrons therefore giving a
low-energy {\it coherent} polaronic band in the spinless case. It is however
easy to show that this form of the spectral function is not compatible with a
$k$-independent self-energy. The self-consistency condition
(\ref{eq:Gp_spinless}) gives rise to a non-zero damping at the Fermi level even in the spinless case. However,
when $\ad $ becomes larger $\alpha$ gets smaller reproducing the anti-adiabatic
coherent behavior at low energy in the spinless and the negative-$U$ behavior
in the spinful cases.

In the anti-adiabatic regime the HLFA approach gives an estimate of the MIT
\beq
\label{eq:MIT_LF}
\lambda_{MIT}=|U/D|_{MIT}\exp(-\alpha^2)
\eeq
where $|U/D|_{MIT} \simeq 2.94$ is the MIT value of the negative-$U$ Hubbard model
\cite{Uc1Uc2neg}.
The pairing MIT occurs for smaller $\lambda$
as the adiabatic regime is approached (see Fig. \ref{fig:PD}).

The phonon PDF can be easily derived within LF approach.
Being the local electron densities parametric variables in the anti-adiabatic
we find
\beq
P(x)=\sum_l w_l P_0(x-x_l)
\eeq
where $w_l$ is the probability of having an occupancy $n_l$,
$x_l$ the relative displacements and $P_0(x)$ the ground state PDF of an
harmonic oscillator. $P_0(x-x_l)$ is then the conditional probability of having a
displacement $x$ given a definite occupation $n_l$.
In the spinless case $n_l=0,1$ with equal probability
giving
\beq
\label{eq:PXLFspinless}
P(X)=\frac{1}{2} \left ( P_0(x-x_0)+ P_0(x+x_0) \right)
\eeq
where $x_0=\sqrt{\lambda}/2$.
A definite polarization can be associated to the ground state if the PDF
becomes bimodal. By requiring $d P(X)/d X\vert_{X=0} > 0$, which simply means
$X=0$ turns from a maximum to a minimum,
we get the usual anti-adiabatic condition for the existence of a polaronic state, i.e.,
$\alpha^2>1$ (see fig. \ref{fig:PD}).

In the spinful case $n_l=0,1,2$
\beq
\label{eq:PXLFspinful}
P(X)= n_d (P_0(X-2X_0)+P_0(X+2X_0))+(1-2n_d)P_0(X)
\eeq
where $n_d= \langle n_\uparrow n_\downarrow\rangle$ is the site double occupancy.
It is worth noting that in  the insulating state $n_d\simeq1/2$,
and the zero-displacement PDF associated to singly occupied sites is depleted.
The existence of a definite polarization is
now associated to a bipolaronic state.
The condition under which  (\ref{eq:PXLFspinful}) becomes bimodal, is 
\beq
exp(-2\alpha^2)(4\alpha^2-1) \ge \frac{1-2n_d}{2n_d}.
\eeq

An estimate for the bipolaronic transition can be obtained by taking $n_d = 1/2$,
which gives $\lambda_{pol}=\gamma/2$. The presence of a fraction of
singly occupied states increase the critical value of $\lambda$  (see Fig. \ref{fig:PD}).
We notice that the spinful PDF for $n_d = 1/2$  maps onto the spinless one
after the {\it same rescaling} of the adiabatic regime (\ref{scaling}).

\subsection{Results from DMFT-ED}
\label{sec:HHFEDresults}
\begin{figure}[htbp]
\begin{center}
\includegraphics[scale=0.4,angle=270]{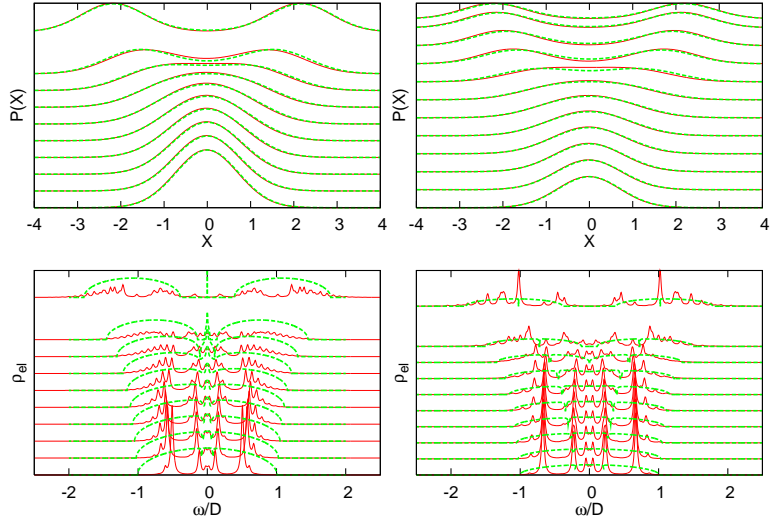}
\end{center}
\caption{DMFT data in the adiabatic regime $\ad=0.1$ spinless
(panels on the left) and
$\ad=0.2$ spinful (panels on the right). The various curves 
refer to different value of
$\lambda$ spanning from  $0.1$ to $1.8$ in the spinless case
and from $0.05$ to $1.1$  in the spinful case
and are shifted according $\lambda$ value.
Upper panels show the phonon PDF
while lower panels the  electronic DOS.}
\label{fig:data_adiab}
\end{figure}

\begin{figure}[htbp]
\begin{center}
\includegraphics[scale=0.4,angle=270]{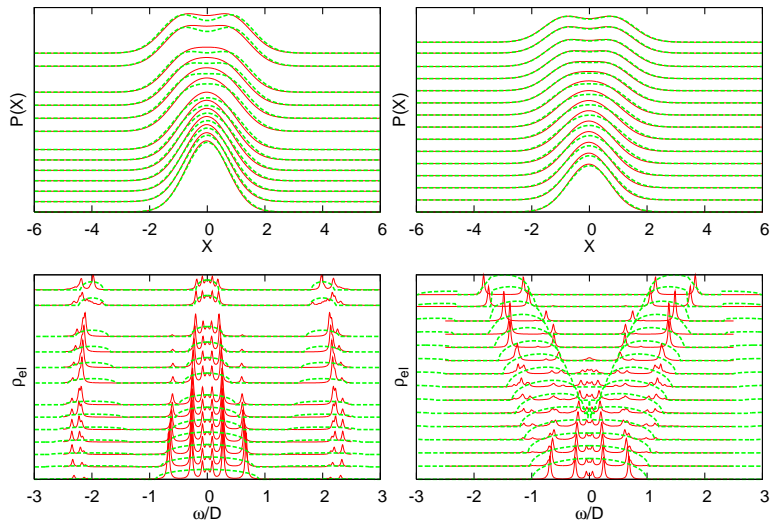}
\end{center}
\caption{DMFT data in the antiadiabatic regime $\ad=2.0$ spinless
(panels on the left) and
$\ad=4.0$ spinful (panels on the right).
The various curves refer to different value of
$\lambda$ spanning from  $0.4$ to $6.5$ in the spinless case
and from $0.2$ to $3.0$  in the spinful case
and are shifted according $\lambda$ value.
Upper panels show the phonon PDF
while lower panels the  electronic DOS.}
\label{fig:data_antiad}
\end{figure}

The behavior of $P(x)$ and $\rho$ obtained from DMFT-ED compared with that of
the previous theories is shown in Figs. \ref{fig:data_adiab}  and
\ref{fig:data_antiad}  for adiabatic and antiadiabatic regime respectively.
In each diagram the value of the quantity shown has been shifted upward
according to the value of the coupling lambda. The values of $\gamma$
and $\lambda$ have been chosen according to the scaling (\ref{scaling}).

Let us first discuss the adiabatic regime.
Polaron crossover is seen as a qualitative change in the shape of phonon PDF 
shown in the upper panels of 
Fig. \ref{fig:data_adiab}  where BO approximation and DMFT results are compared.
The anharmonicity due to e-ph interaction increases as the
coupling increases leading first to a non-Gaussian and finally to a bimodal PDF
at $\lambda > \lambda_{pol}$.
This behavior signals the appearance of static distortions, even if we are neglecting
any ordering between them.
>From Fig. \ref{fig:data_adiab} is evident that BO approximation works well in
{\it both} the metallic and the polaronic regimes. The reason for the accuracy of the
BO procedure in the polaronic regime is that, contrary to its usual
implementation in the weak e-ph coupling\cite{russianBO},
here we  take into account the anaharmonicity
through Eq. (\ref{eq:BO}) in a non perturbative way.
However, BO does not accurately reproduce the  phonon PDF around the polaron crossover.
In this case electron and phonon states are  strongly entangled, and cannot be approximated
properly by a disentangled BO state.
By a comparison of the spinless and spinful cases in Fig. \ref{fig:data_adiab} we
see that the occurrence of the MIT does not influence much the differences between
full DMFT and BO, which are in both cases relevant near the polaron crossover.
In the lower panels of Fig. \ref{fig:data_adiab} we compare the electronic
DOS from ED-DMFT with  IBOCPA. The different behavior of the spinless and
spinful case is not so evident.
However, comparing the spinless spectrum with the corresponding
spinful, we see that a quasiparticle peak is present in the former case,
while a depletion of low energy part is much more evident in the latter.
At strong coupling the discretization of the bath
inherent to the ED solution of DMFT
does  not allow us to identify  a well defined quasiparticle peak in the spinless case.
A more careful analysis of the quasi particle spectral weight \cite{Max1}
shows however that no pairing MIT occurs in the spinless case.
Notice that the IBOCPA seems to be much closer to DMFT-ED in the spinless than in
the spinful case where the CPA approximation for the electronic degree of
freedom is apparently much less adequate.

The upper panel of Fig. \ref{fig:data_antiad} shows that 
 the  phonon PDF becomes bimodal at a very large value of
the coupling. HLFA  overestimates polaronicity in the spinless case while it
behaves better in the spinful case.
In the lower panels of Fig. \ref{fig:data_antiad} the electron DOS is compared
with the results of HLFA in the antiadiabatic regime. The different behavior of
spinless and spinful cases is marked here by the presence of a pairing 
MIT in the former, well before the 
(bi)polaron crossover. HLFA correctly catches the gross behavior of the DOS.
Notice that at the strong coupling the CPA employed to obtain the lower
diagrams accurately reproduces both the position and the width 
of the side bands.

The different behavior of spinless and spinful system can be easily understood
in terms of strong coupling anti-adiabatic perturbation theory for the original
lattice problem \cite{Freericks-strong} introducing the charge sector
pseudo-spins \cite{Micnas}. 
In the spinful case at second order in hybridization $V_\kvec$ 
an anisotropic Kondo Hamiltonian
can be obtained \cite{Cornaglia-condmat,StJ,PCarta}.
%\begin{eqnarray}
%\label{eq:KondoHIM}
%H &=& H^{\prime}+J_{\parallel}\sum_\kvec v_k\rho^z_f \rho^z_c(\kvec)+
%\nonumber\\
%&+&\frac{J_{\perp}}{2}\sum_\kvec v_k (\rho^+_f \rho^-_c(\kvec) + \rho^-_f \rho^+_c(\kvec))
%\end{eqnarray}
%where $H^\prime$ contains all the terms of  (\ref{eq:AndersonLF}) which are not
%proportional to $V_\kvec$,
%$V^2=\sum_k|V_\kvec|^2$ and $|v_\kvec|^2=|V_\kvec|^2/V^2$.
%$\rho$ are pseudo-spin operators corresponding to bath ($\rho_c$) or to
%impurity ($\rho_f$) states,
The anisotropic
Kondo couplings (see also Eq. (8) of Ref.\cite{Cornaglia}) reads
\beq
J_{\parallel,\perp} =
\frac{8V^2}{D}\sum_m (\pm)^m
\frac{e^{-\alpha^2}\alpha^{2m}}{m!(m\ad+\lambda/2)},
\eeq
where the $+$ ($-$) sign is taken for the $\parallel$ ($\perp$)
coupling and we have assumed for simplicity $V=\sum_k V_k$.
In the spinless case the processes
leading to $J_{\perp}$ do not exist while the remaining $J_{\parallel}$ is
solely associated to charge fluctuations. 
%i.e. 
%it will correspond to a
%$f$-charge $c$-charge interaction in the effective Hamiltonian
%\beq
%H = H^\prime+J_{\parallel}\sum_k v_k\rho^z_f \rho^z_c(\kvec)
%\eeq
%where now $\rho^z_f=f^\dagger f-1/2$.
As opposed to the charge Kondo effect of the spinful case, no Kondo effect is expected 
for spinless fermions.
A strong coupling estimates of $J_{\parallel}$ gives
\beq
J_{\parallel}\simeq \frac{8V^2}{D\lambda}
\eeq
and $J_{\perp}/J_{\parallel} \propto \exp(-2\alpha^2)$ which means an
exponential suppression of the superconductivity versus charge
correlation at strong coupling due to retardation effects
\cite{Cornaglia-condmat,StJ}. In the anti-adiabatic
limit $\ad\rightarrow \infty$ instead the Kondo couplings becomes isotropic.

\begin{figure}[htbp]
\begin{center}
\includegraphics[scale=0.33,angle=270]{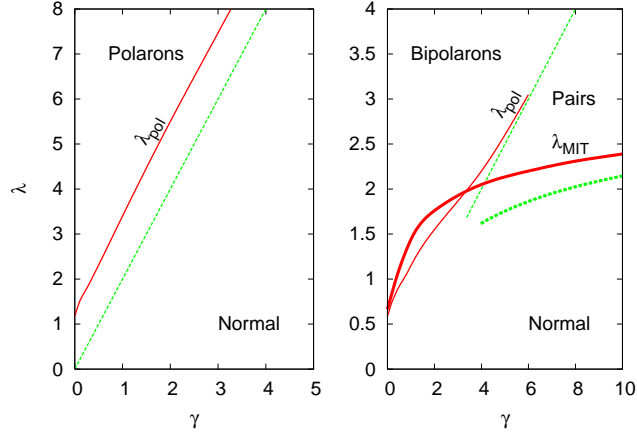}
\end{center}
\caption{Phase diagrams of the spinless (left) and spinful (right)
$T=0$ Holstein model at half filling. 
Left panel: the bold line is the polaron crossover from bimodality
of $P(x)$ and the dotted line is the anti-adiabatic estimate
$\lambda_{pol}=2\gamma$ for the polaron crossover.
Right panel:bold curve is the bipolaronic MIT from vanishing of quasi-particle
spectral weight $Z$,
thin solid line the polaron crossover, bold dotted line is the anti-adiabatic
prediction for bipolaronic MIT (Eq. (\ref{eq:MIT_LF})),  light dotted line is
the anti-adiabatic estimate $\lambda > \gamma /2$ for the polaron crossover.}
\label{fig:PD}
\end{figure}

The above observations  can be summarized in the phase diagrams of Fig.
\ref{fig:PD}.
The polaron crossover line $\lambda_{pol}$ is 
strongly $\ad$ dependent in both cases.
Above this line a polaronic (bipolaronic) regime is attained in the spinless (spinful) case.
In the spinful case, we also have a $\lambda_{MIT}$ line, which separates a 
normal phase from a paired insulating phase\cite{Max1}. 
For large phonon frequency we can have pairs without
bipolaronic behavior, as it can be understood by recalling that in the 
antiadiabatic limit the Holstein model becomes an attractive Hubbard model,
where no polarization is associated to the pairing.

%figures:
%/home/ciuk/LAVORI/SEMINARS/Varenna05/seminar/fig_DOSvsBO.eps
%/home/ciuk/LAVORI/SEMINARS/Varenna05/seminar/fig_ZOm_w0=0.1.eps
%/home/ciuk/LAVORI/SEMINARS/Varenna05/seminar/fig_ZOm_w0=0.1_s.eps
%/home/ciuk/LAVORI/SEMINARS/Varenna05/seminar/fig_DOSvsLF.eps
%/home/ciuk/LAVORI/SEMINARS/Varenna05/seminar/fig_ZOm_w0=1.0.eps
%/home/ciuk/LAVORI/SEMINARS/Varenna05/seminar/fig_ZOm_w0=1.0_s.eps
%/home/ciuk/LAVORI/SEMINARS/Varenna05/seminar/fig_PD.eps

\section{Half filled Holstein model: spinless vs spinful cases at $T>0$}
\label{sec:HHFQMC}
Using the QMC procedure described in section \ref{sec:DMFT-QMC} we are able to
study the normal phase at finite temperature. At fairly high temperature the MIT becomes a
crossover and we are faced with the problem of finding a 
suitable quantity to locate unambiguously this crossover.
In analogy with the phonon PDF used to mark the polaron crossover 
we can define a distribution of a quantity
that locates the pairing crossover. Let us define the distribution 
of the center of mass $X_c$ (``centroid'') of the phonon path
\beq
\label{PXc}
P(X_c)= \left \langle \delta(X_c-\frac{1}{\beta}\int_0^\beta x(\tau)d\tau) \right \rangle
\eeq
where the averages are evaluated over the action
 (\ref{Sel}-\ref{Selph}). In the same formalism the phonon PDF is
 the distribution of the the endpoint $x=x(0)=x(\beta)$.

The meaning of the centroid variable $X_C$ has been discussed in 
\cite{Kleinert} for a single particle in a binding potential.
Here the variable $X$ represents the fluctuating position of the particle, and
 $X_c$ is the classical position of the  particle \cite{Kleinert}. 
For an heavy particle, the classical limit holds, so that 
$P(X)$ and $P(X_C)$ coincide \cite{Kleinert}.
The lighter is the particle, the broader the wave function, increasing
the variance of $P(X)$ while $P(X_c)$ turns out to be essentially determined by
the binding range of the potential.
Here, we use $P(X_C$) for the many-body problem, and propose that pairing
can be associated with a multimodal behavior in 
$P(X_c)$ which takes place at a given value of the coupling $\lambda_{pair}$.
\cite{centroide} 
The ability of our estimator to determine the pairing crossover can be 
understood by inspecting the interaction term (\ref{Selph}).
In the adiabatic limit ($\ad \rightarrow 0$) the kinetic term 
forces the phonon path to be $\tau$-independent and the phonon field becomes 
classical.
In this limit $X_c$ is equal to $X$ and $P(X)$ and 
$P(X_c)$  obviously coincide. Thus the centroid distribution becomes bimodal 
when the system is polarized, i.e., $\lambda_{pair}=\lambda_{pol}$, which is 
exactly what one expects 
since a static field can induce pairing only with a finite polarization.
Notice that in this sense the bimodality of $P(X_c)$ is a {\it precursor} of 
the
actual pairing MIT which occurs at $T=0$ and $\omega_0=0$ at a {\it larger}
value of the coupling \cite{Millis-adiab} (see fig. \ref{fig:PD}).

On the other hand, in the opposite atomic ($D \rightarrow 0$, 
$\ad \rightarrow \infty $)
limit the electron density becomes a constant of motion. 
Therefore Eq. (\ref{Selph}) takes the transparent form 
$-\sqrt{\lambda} (n-1) \int_0^\beta d\tau x(\tau)$ where the electron density is 
directly coupled to the centroid $X_c$. 
The average appearing in Eq. (\ref{PXc}) is readily carried out, giving
\beq
\label{PXcAtomic}
P(X_c) \propto \exp \left[-\beta (\frac{X_c^2}{2}-\frac{1}{\beta}\log (2
\cosh (\beta\sqrt{\lambda}X_c+1))) \right]
\eeq
which becomes bimodal for $\lambda>\lambda_{pair}=2T$. This is exactly the scale where double
occupancies start to proliferate in the atomic limit. 
Therefore  the bimodality of $P(X_c)$ 
correctly signals the onset of pairing also in the antiadiabatic regime. 
In the same limit, it can be proved that the endpoint distribution $P(X)$ has
a variance which scales with $1/\sqrt{\Delta \tau}$ 
and as a consequence no definite polarization may occur.
We finally notice that the $D \rightarrow 0$ 
limit of adiabatic $P(x)$ \cite{Millis-adiab}
coincides with $P(X_c)$ of Eq. (\ref{PXcAtomic}).
Since for  $\omega_0=0$ the distributions of $X$ and $X_c$ coincide, 
we conclude that in the atomic limit $P(X_c)$ 
is the same for $\omega_0=0$ and 
$\omega_0=\infty$. This suggests that the pairing crossover may depend
on $\omega_0$ more weakly than the polarization one.

\subsection{Results from DMFT-QMC}
\begin{figure}[htbp]
\begin{center}
\includegraphics[width=6.5cm,angle=270]{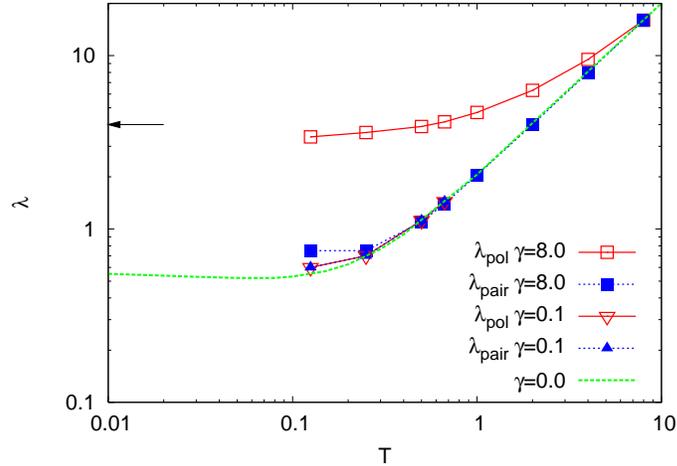}
\end{center}
\caption{$\lambda_{pair}$  and 
$\lambda_{pol}$ at $\ad=8$. 
The  dashed line represents the $\ad = 0$ result where 
$\lambda_{pair}=\lambda_{pol}$. The dashed arrow indicates 
the zero-temperature results for the polaron crossover $\lambda_{pol}$ for $\gamma=8$.}
\label{PD}
\end{figure}

To analyze the evolution of $P(X)$ and $P(X_c)$ at finite $D$ and $\omega_0$ 
we use DMFT-QMC. The numerically exact results, shown in Fig.\ref{PD}, 
clearly show that $P(X)$ and $P(X_c)$ tend to coincide in the 
relatively adiabatic case $\ad = 0.1$, as expected from the previous
arguments about the adiabatic limit. The two quantities are clearly 
 different for $\ad = 1$ and $8$. 
For temperatures smaller than $\omega_0$, the polarization
crossover $\lambda_{pol}$ moves to larger
values as $\ad $ is increased, while  the line ($\lambda_{pair}$) where $P(X_c)$
becomes bimodal is only slightly shifted to larger couplings with increasing
$\ad$. This is strongly reminiscent of the behavior of the metal-insulator
transition in the Holstein model at $T=0$, whose critical coupling
is slowly increasing with $\ad$ and then saturates to the asymptotic
$\ad=\infty$ value\cite{Max1}. The polarization
crossover is instead roughly proportional to $\ad$\cite{Max1}.
Both at zero and at finite temperature the line where the centroid becomes
bimodal does not coincide with the metal-insulator line, but it can be
considered a precursor which depends in a very similar way on $\ad$ 
and on $\lambda$.

Our DMFT results can also be compared with the semi-analytical results for
$\ad=0$ \cite{Millis-adiab} (thin dashed line 
in Fig.\ref{PD}). The $\ad=0.1$ case is in very good
agreement with the adiabatic result, and also the $\ad=1$ and $8$ cases,
at high temperature, fall on the same curve. 
The centroid distribution depends weakly on $\ad$, as suggested by the atomic limit. 
Interestingly, the adiabatic result displays a re-entrance at low temperatures as a non monotonic behavior
of $\lambda_{pair}$ and $\lambda_{pol}$ as a function of temperature.
Although the QMC simulations do not reach sufficiently low temperatures, we
find that the re-entrance is present also for $\ad$ different from zero,
as indicated by the arrows in Fig.\ref{PD}, which mark $T=0$ results for the 
polarization crossover for the Holstein model\cite{Max1}.

\section{Conclusions}
The normal state properties of strongly coupled electron phonon systems  can be
drawn from the comparison of electronic spectral and phonon PDF properties.

Qualitative changes in phonon PDF signals a polaronic crossover while an
electronic MIT can be seen from a gap in the electronic DOS or by the vanishing of the quasi particle
spectral weight. This last transition can be observed only at sufficiently high density provided the
Coulomb repulsion is neglected. In the limit of low density the carrier show a polaronic behavior 
trough developing a definite polarization in the phonon PDF while at large densities occupied and empty 
sites makes the phonon PDF bimodal. At the polaron crossover fluctuation of phonon coordinates tends to be
larger than those at any other coupling. In this regime Born-Oppenheimer approximation is shown to fail
also when phonon frequency is much less then electron bandwidth. For a single hole in the $t-J$ model a
further source of localization is due to magnetic superexchange energy which tends to localize the
spin-defect spreading due to the presence of the hole. 

At non zero temperature the pairing MIT becomes a crossover. To locate the point of this crossover it is
possible to define a suitable distribution associated to the phonon classical position which is coupled to
the electron density in the non-adiabatic regime. 

There are several limitations in the DMFT method used to obtain the present results. 
The biggest one is the non dispersive nature of bosonic excitation. In the $t-J$ case the absence of
spin-wave dispersion lead to localization of the single hole. However local properties seem to be quite
well represented by DMFT in when compared with those found at finite dimensions \cite{MishchenkoVarenna}.
In the half filled Holstein model case bipolarons do not have coherent motion therefore the vanishing of the
single particle spectral weight implies a MIT.
Several extensions of DMFT such as Extended-DMFT for the $t-J$ model 
\cite{EDMFT} or CDMFT for the Hubbard model \cite{CDMFT}
try to overcome such limitations. Work along these lines for the Holstein and Holstein-Hubbard model
are currently in progress.


\begin{thebibliography}{0}
\bibitem{NoteKF2} This expression is equivalent to the CPA free energy of the Kimball Falikov model \cite{BrandtMielsch,ChungFreericks}
\bibitem{NotaIBOCPA} Notice that in contrast with  ref.\cite{PCarta} here iteration to adjust TSMP parameter is fully implemented.
\end{thebibliography}
\end{document}